\documentclass[10pt,aps,prd,nofootinbib,superscriptaddress,twocolumn,preprintnumbers,balancelastpage]{revtex4}
\usepackage{graphicx}
\usepackage{epstopdf}
\usepackage{bm}

\usepackage[colorlinks]{hyperref}
\hypersetup{
     colorlinks   = true,
     citecolor    = red,
 linkcolor=blue
}
\usepackage{color}
\usepackage{soul}
\usepackage{amsmath}
\usepackage{amssymb}
\usepackage{mathtools}
\usepackage{xfrac}
\usepackage{multirow}
\usepackage{placeins}

\newcommand{\gaga}{{\gamma\gamma}}
\newcommand{\MET}{{\slash\!\!\!\!E}_{T}}
\newcommand{\vMET}{{\vec{\slash\!\!\!\!E}_{T}}}

\newcommand{\ext}{{\rm ext}}
\newcommand{\trig}{{\rm trig}}
\newcommand{\fid}{{\rm fid}}
\newcommand{\reco}{{\rm reco}}

\usepackage{feynmp-auto}
\usepackage[usenames,dvipsnames,svgnames]{xcolor}
\definecolor{Purple}{rgb}{0.9,0.2,0.5}

\begin{document}

\newcommand{\toname}{{\sqrt{s^{c}_{\gaga}}}}

\newcommand{\ion}{{\textrm{ion}}}
\newcommand{\Nu}{{\textrm{nu}}}
\newcommand{\GeV}{{\rm GeV}}
\newcommand{\TeV}{{\rm TeV}}
\newcommand{\fm}{{\rm fm}}
\renewcommand{\max}{{\rm max}}
\newcommand{\cm}{{\rm cm}}

\graphicspath{{Figures/}}

\title{
Tracking down Quirks at the Large Hadron Collider  
}

\author{Simon Knapen}
\author{Hou Keong Lou}
\author{Michele Papucci}
\affiliation{Department of Physics, University of California, Berkeley, California 94720, USA}
\affiliation{Theoretical Physics Group, Lawrence Berkeley National Laboratory, Berkeley, California 94720, USA}
\author{Jack Setford}
\affiliation{Department of Physics and Astronomy, University of Sussex, England, UK}


\begin{abstract}
Non-helical tracks are the smoking gun signature of charged and/or colored quirks, which are pairs of particles bound by a new, long-range confining force. We propose a method to efficiently search for these non-helical tracks at the LHC, without the need to fit their trajectories. We show that the hits corresponding to quirky trajectories can be selected efficiently by searching for co-planar hits in the inner layers of the ATLAS and CMS trackers, even in the presence of on average 50 pile-up vertices. We further argue that backgrounds from photon conversions and unassociated pile-up  hits can be removed almost entirely, while maintaining a signal reconstruction efficiency as high as $\sim$70\%. With the $300\, \text{fb}^{-1}$ dataset, this implies a discovery potential for string tension between 100 eV and 30 keV, and colored (electroweak charged) quirks as heavy as 1600 (650) GeV may be discovered. 
 
%
\end{abstract}

\date{\today}

\maketitle

\section{Introduction}
\label{sec:intro}
With run II of the Large Hadron Collider (LHC) well underway, signatures of beyond the Standard Model physics have yet to reveal themselves. As the LHC transitions to its luminosity driven-phase, its focus will shift toward precision measurements and low rate signals. It is hereby imperative to consider new physics signatures that  may not yet be covered; a task which has become increasingly difficult as the collaborations have greatly expanded and refined their search strategies in recent years. Nevertheless, there is considerable room for further progress, in particular in the context of long-lived exotica. The reason is that triggering and tracking often raise unique challenges, such that the sensitivity of more traditional searches is very poor or non-existent, and specialized strategies are needed. Nonetheless, once these challenges are addressed, these dedicated exotica searches (e.g.~Long-lived particles, R-hadrons, disappearing tracks, hidden valleys \cite{Drees:1990yw, Hewett:2004nw, Arvanitaki:2005nq,Cirelli:2005uq,Strassler:2006im}), have resulted in some of the most stringent experimental limits to date \cite{Mehlhase:2016vuh, Khachatryan:2016sfv, ATLAS-CONF-2016-103,CMS:2014wda, ATLAS-CONF-2017-017,CMS:2014gxa}, precisely because of their qualitative departure from known standard model phenomena.

%
%

In this paper, we consider the quirks scenario  \cite{Kang:2008ea}, for which traditional tracking algorithms break down. A quirk/anti-quirk pair is a pair of new heavy stable charged particles (HSCP's), that is connected by a flux tube of dark gluons. Such quirks can be present in models of dark matter \cite{Kribs:2009fy} or neutral naturalness, like the quirky little Higgs \cite{Cai:2008au}, folded supersymmetry \cite{Burdman:2006tz,Burdman:2008ek} and certain twin Higgs models \cite{Craig:2015pha,Craig:2016kue}.
The regime we consider here is defined by a large hierarchy between the quirk mass ($m_Q$) and the dark confining scale $\Lambda$, {\it i.e.~}$m_Q\gg\Lambda$. In this limit, the breaking of the dark flux tube, by pulling a quirk/anti-quirk pair from the vacuum, is suppressed by $\sim\exp(-m_Q^2/\Lambda^2)$. This is to be contrasted with standard model QCD, for which $m_Q \ll \Lambda$. In QCD, an excited flux tube can therefore easily break into multiple bound states, which is the process known as hadronization. For quirks, the flux tube does not break and instead induces a spectacular, macroscopic oscillatory motion before the quirks eventually annihilate. 
In the center of mass (CM) frame of the quirk/anti-quirk pair, the characteristic amplitude of this oscillation is
\begin{equation}
d_{\cm}\sim 2 \;{\rm cm}\,\left(\gamma -1 \right)\bigg(\frac{m_Q}{100\; {\rm GeV}}\bigg)\bigg(\frac{{\rm keV}}{\Lambda}\bigg)^2\,,
\label{eq:d_cm}
\end{equation}
where $\gamma=1/\sqrt{1-v^2}$ is the Lorentz boost factor of quirks at the moment of their production. 

For large $\Lambda \gtrsim 30$ keV, the oscillation length will typically be smaller than the detector resolution (roughly $\sim 100\; \mu {\rm m}$), and the combined motion of the quirks is resolved as a single, nearly straight track. In the track reconstruction, this would be seen as a very high $p_T$ track with high $dE/dx$.  A dedicated search of this type was carried out by the D0 collaboration at the Tevatron \cite{Abazov:2010yb}. This search has not yet been repeated at the LHC, but it is conceivable that the existing HSCP searches have nevertheless sensitivity to this scenario. We leave this possibility for future work. In the opposite regime, where $\Lambda \lesssim 100$ eV, the length of the string is of the order of the detector size or larger. For this regime it has recently been shown that the existing HSCP searches already set rather strong limits \cite{Farina:2017cts}. 

In the intermediate regime where $100\, \mathrm{eV} \lesssim\Lambda \lesssim 10$ keV, most events will have an oscillation amplitude of roughly $d \sim 0.1$ to $10$ cm. In this case, no tracks are reconstructed with existing algorithms, and the only current constraint comes from the jets + $\MET$ search \cite{Farina:2017cts}.   Although cm-size oscillating tracks would be a truly spectacular signature, it is thought to be very difficult to design a reconstruction algorithm for such tracks, especially with current high pile-up conditions and given that the $m_Q$ and $\Lambda$ are not a priori known. Even for fixed $\Lambda$ and $m_Q$, the trajectories depend strongly on the initial velocities of the quirks and can differ greatly on an event-by-event basis.  

\begin{figure}[t]
\includegraphics[trim={.5cm 1.3cm 1.3cm .5cm},clip,width=0.4\textwidth]{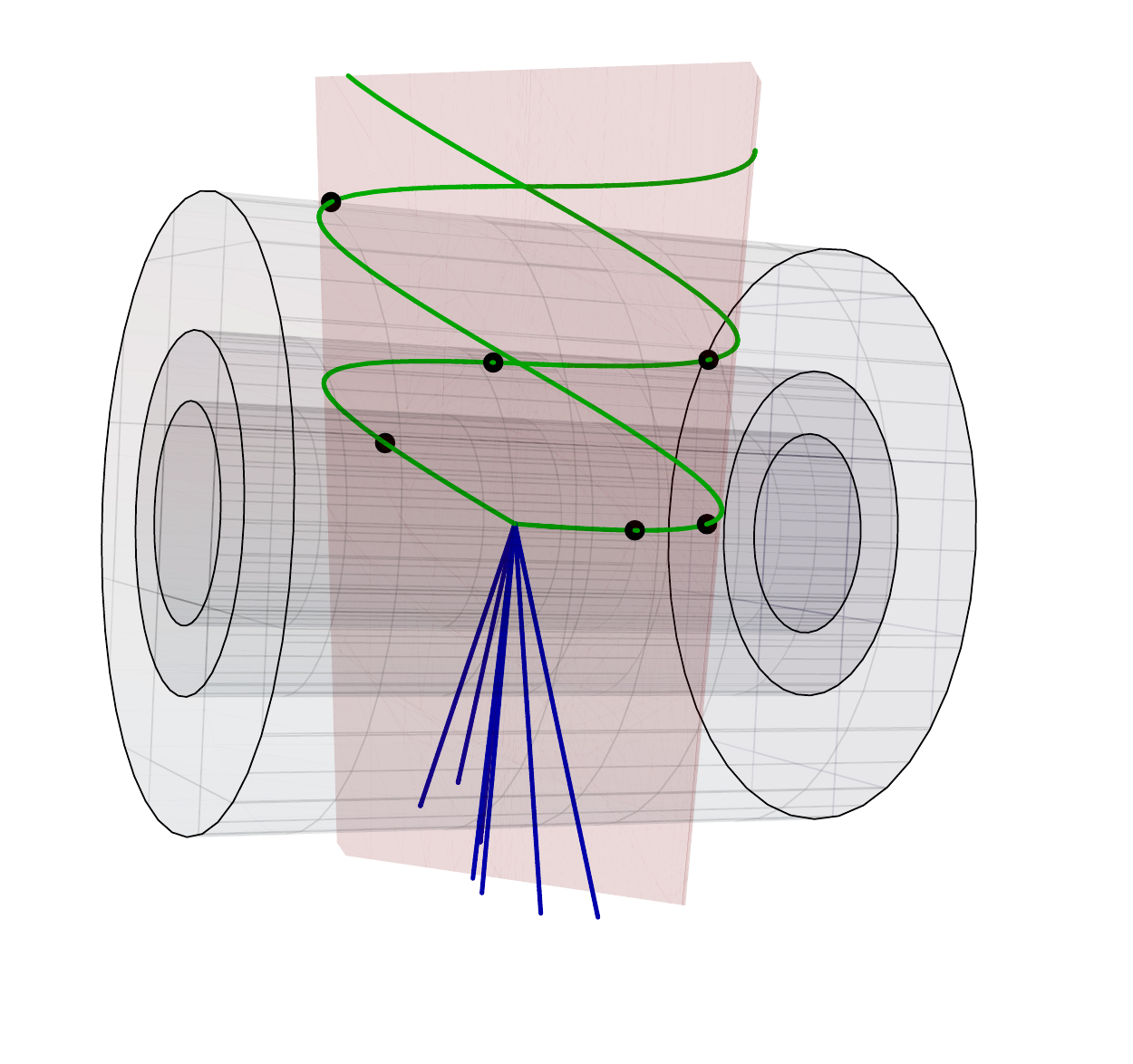}
\caption{Schematic event display of a pair of quirks (green) with an ISR jet (blue). The cylinders represent the three innermost layers of the ATLAS/CMS tracker. The hits (black dots) all lie on a single plane (shaded red). \label{eventdisplay}}
\end{figure}

Rather than attempting to reconstruct the tracks directly, we will therefore take advantage of some of the universal features of the motion of two particles subject to a central force. This allows us to develop a strategy that is largely independent of $\Lambda$, $m_Q$ and the kinematic configuration of the event. In particular, we will argue that the angular momentum of the quirk/anti-quirk system is approximately conserved as it traverses the \mbox{ATLAS}/CMS tracker. Since the quirk/anti-quirk system is initially produced with negligible angular momentum, \emph{the trajectories lie on a plane} to a good approximation. The idea is therefore to search for pairs of hits in each layer which all lie on a single plane (See Fig.~\ref{eventdisplay}). 

The remainder of this paper is organized as follows: In Sec.~\ref{sec:dynamics}, we review quirk dynamics and how to model their motions. We present details on our search strategy in Sec.~\ref{sec:strategy} and the main results and sensitivity estimates in Sec.~\ref{sec:results}. We reserve some additional results on $dE/dx$ for App.~\ref{app:dedx}.

\section{Quirk Dynamics}
\label{sec:dynamics}
At the LHC, quirks can be pair-produced through either electroweak (Drell-Yan) and/or QCD interactions. Below we study the dynamics of quirks after they are pair-produced. As our benchmarks scenarios, we will consider vector-like quirks in the $(1,1)_1$ and $(3,1)_{\frac{2}{3}}$ representations. In the latter case, the quirks will quickly hadronize into quirk-hadrons, and the probability for those final states to have $\pm 1$ charges is roughly 30\% as estimated using \texttt{Pythia8} \cite{Sjostrand:2014zea}. Our analysis is largely independent on the charges of the quirk-hadrons, as long as both quirk-hadrons carry non-zero electric charge, such that they leave a signal in the inner trackers of ATLAS and CMS. In what follows we will loosely refer to the quirk-hadrons as quirks.

The quirks are approximately free right after they are produced. As their separation length becomes larger than $\Lambda^{-1}$, confinement will lead to an unbreakable flux-tube connecting the two quirks. This system can be described by the Nambu-Goto action with massive endpoints, which has been shown to correctly capture the properties of the heavy quark potential in QCD \cite{Luscher:2002qv}. More general actions are possible, but should not affect our results significantly, as long as the string tension is much larger than the Lorentz force exerted by the magnetic field. The action for the quirks and the flux-tube (effectively a string) is then,
\begin{equation} 
S = - m_Q \sum_{i=1,2} \int d\tau_i -\Lambda^2 \int d A + S_{\ext}\,,
\label{eq:nambugoto}
\end{equation}
where $A$ is the area of the string worldsheet, $\tau_i$ the proper time of the two quirks, and $S_{\ext}$ describes external forces on the system. The boundaries of the string worldsheet are fixed to be the worldlines of the quirks. Note that we have taken $\Lambda^2$ to be the string tension, which will also serve as a precise definition for $\Lambda$. Eq.~\ref{eq:nambugoto} leads to the following sets of equations for the quirks
\begin{equation}
\frac{\partial}{\partial t}(m\gamma {\bm v}) = -\Lambda^2\left( 
\frac{\sqrt{1- {v}_\perp^2}}{v_\parallel}\,{{\bm v}_\parallel} + \frac{v_\parallel}{\sqrt{1 - {v}_\perp^2}} {\bm v}_\perp \right) + {\bm F}_\ext,
\label{eq:eom}
\end{equation}
where $\bm v$ is the quirk velocity, $\bm v_\parallel$ and $\bm v_\perp$ are the components of the velocity parallel and perpendicular to the string ($\bm v_\parallel + \bm v_\perp = \bm v$). There is one equation for each quirk, and the dynamics of the string in general leads to another, very complicated partial differential equation that couples to Eq.~\ref{eq:eom}.  Fortunately, in the region where $\Lambda\gg 100$ eV, the force from the string is large compared to other interactions, and the string can be approximated as straight. In this limit, and in the center of mass frame, ${\bm v}_{\parallel}$ will lie along the displacement vector between the quirks, and Eq.~\ref{eq:eom} alone suffices to describe the motion of the quirks. Ignoring ${\bm F}_{\rm ext}$, and for a pair of quirks produced back-to-back with initial velocity $v$, the motion for one period $0 \le t \le 2 v\gamma m_Q/\Lambda^2$ is given by
\begin{align}
  d_{\rm cm}(t)=\frac{m_Q}{\Lambda^2}\left[\gamma - \sqrt{1+\left(\frac{ \Lambda^2t}{m_Q} - v\gamma \right)^2}\, \right]\,,
\label{eq:trajectory}
\end{align}
where $\gamma = 1/\sqrt{1-v^2}$. This gives the amplitude in Eq.~\ref{eq:d_cm}.

In ATLAS and CMS, the trajectory in Eq.~\ref{eq:trajectory} will be modified by the inclusion of ${\bm F}_\ext$, which is the Lorentz force exerted by the magnetic field as well as forces exerted during the passage through the detector material. Then, to justify our proposed search strategy, we must verify two crucial features of the quirk trajectories taking ${\bm F}_\ext$ into account:
\begin{enumerate}
\item The probability that the quirks annihilate before reaching the outer part of the inner tracker is very small.
\item The quirk/anti-quirk system does not pick up a large amount of angular moment as it traverses the detector material and the magnetic field.
\end{enumerate}

\begin{figure}[t]
\includegraphics[trim={1.2cm 3.2cm 2cm .6cm},clip,width=0.45\textwidth]{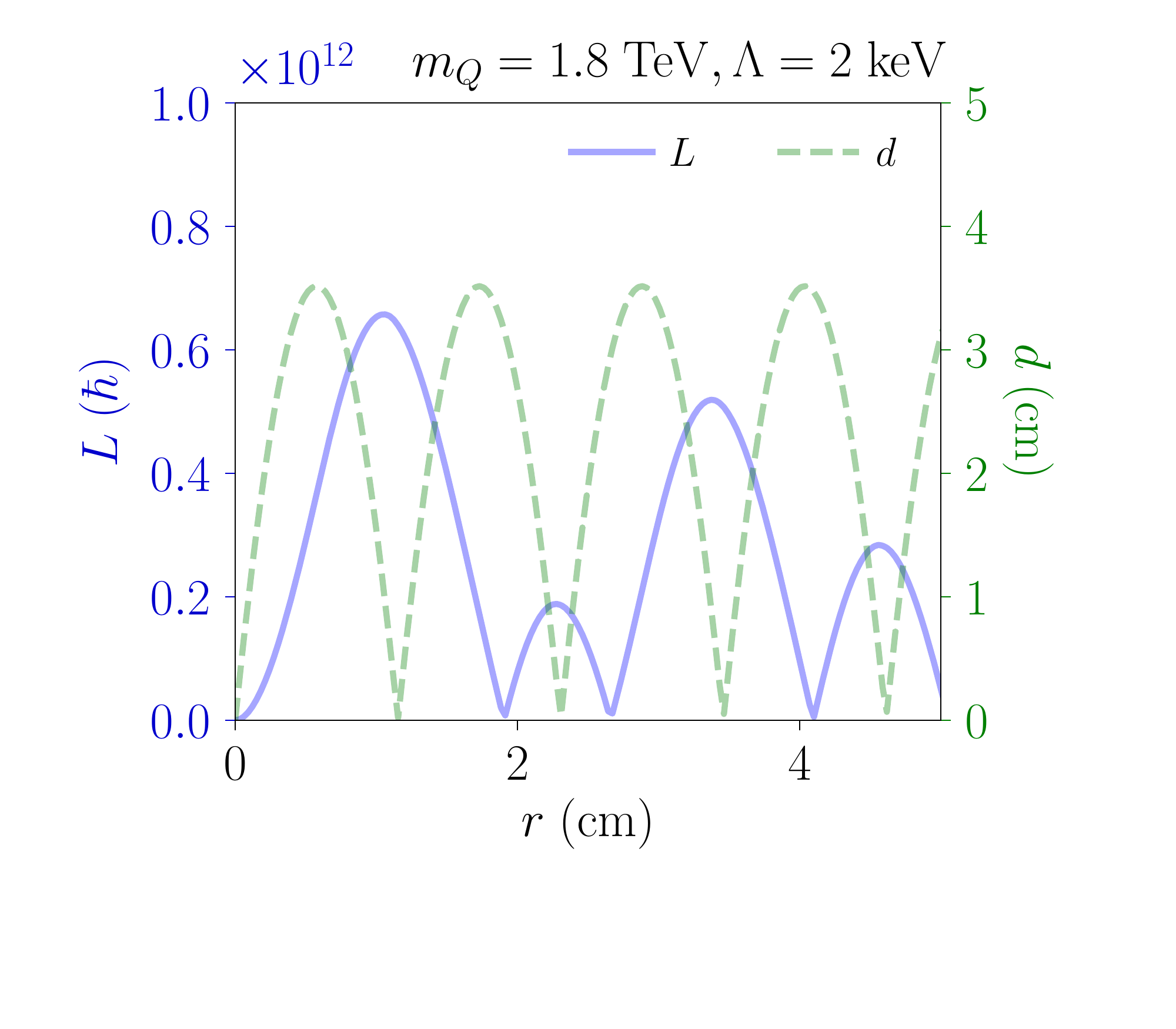}
\caption{Angular momentum $L$ and the relative distance between the quirks $d$, as a function of the radial distance of the center of mass to the interaction point for a representative event, with $B=2$ T. The displacement $d$ varies several orders of magnitude over the quirk trajectories, but despite the appearance on this figure, it does not vanish except at the origin (r=0).}
\label{fig:angular_momentum}
\end{figure}

It is straightforward to see that a typical quirk/anti-quirk system does not annihilate in the presence of a magnetic field, as the $B$-field will induce a macroscopic amount of internal angular momentum in the system, which will prevent it from annihilating. To estimate the effect of the $B$-field, it is useful to move to the center of mass frame. In this frame, the magnetic field is seen as an $E$-field, which introduces a torque
\begin{equation}
{\bm \tau } \sim 2{\bm d} \times (e {\bm E}_{\rm cm}) = 2e\gamma_{\cm} \,{\bm d} \times ({\bm v}_{\rm cm} \times {\bm B}),
\end{equation}
where ${\gamma_\cm=1/\sqrt{1-v_\cm^2}}$ and ${\bm v}_{\rm cm}$ is the center-of-mass velocity. $2{\bm d}$  is the typical displacement of the quirks in the center of mass frame, and ${\bm B}$ the magnetic field in the lab frame. The angular momentum that is picked up in a single oscillation with period $\Delta t$ is roughly
\begin{align}
L&\sim |{\bm \tau }| \Delta t \sim  e v_{\cm}\gamma_\cm v\gamma (\gamma-1)\frac{m_Q^2 B}{\Lambda^4}\\
&\sim  10^{12} \hbar \; \bigg(\frac{v_{\cm}v^3}{0.1}\bigg)\bigg(\frac{2\,\text{keV}}{\Lambda}\bigg)^4\!\bigg(\frac{m_Q}{1.8\,\text{TeV}}\bigg)^2\!\bigg(\frac{B}{2\, \mathrm{T}}\bigg)\,,
\end{align}
where we used $|{\bm d}|\sim(\gamma-1)\frac{m_Q}{\Lambda^2}$, $\Delta t\sim 2v\gamma \frac{m_Q}{\Lambda^2}$ and taken the non-relativistic limit. For such large values of the angular momentum, the annihilation probability is negligible. Equivalently, it is possible to show that the distance of closest approach is much larger than $1/m_Q$. The angular momentum does however oscillate along the trajectory of the quirks. Although whenever $|L|=0$, the separation between the quirks is large, and annihilation is suppressed by a small wave-function overlap. This is illustrated in Fig.~\ref{fig:angular_momentum} for a sample event.

\begin{figure}[t]
\includegraphics[trim={1.2cm 3.2cm 1cm .6cm},clip,width=0.45\textwidth]{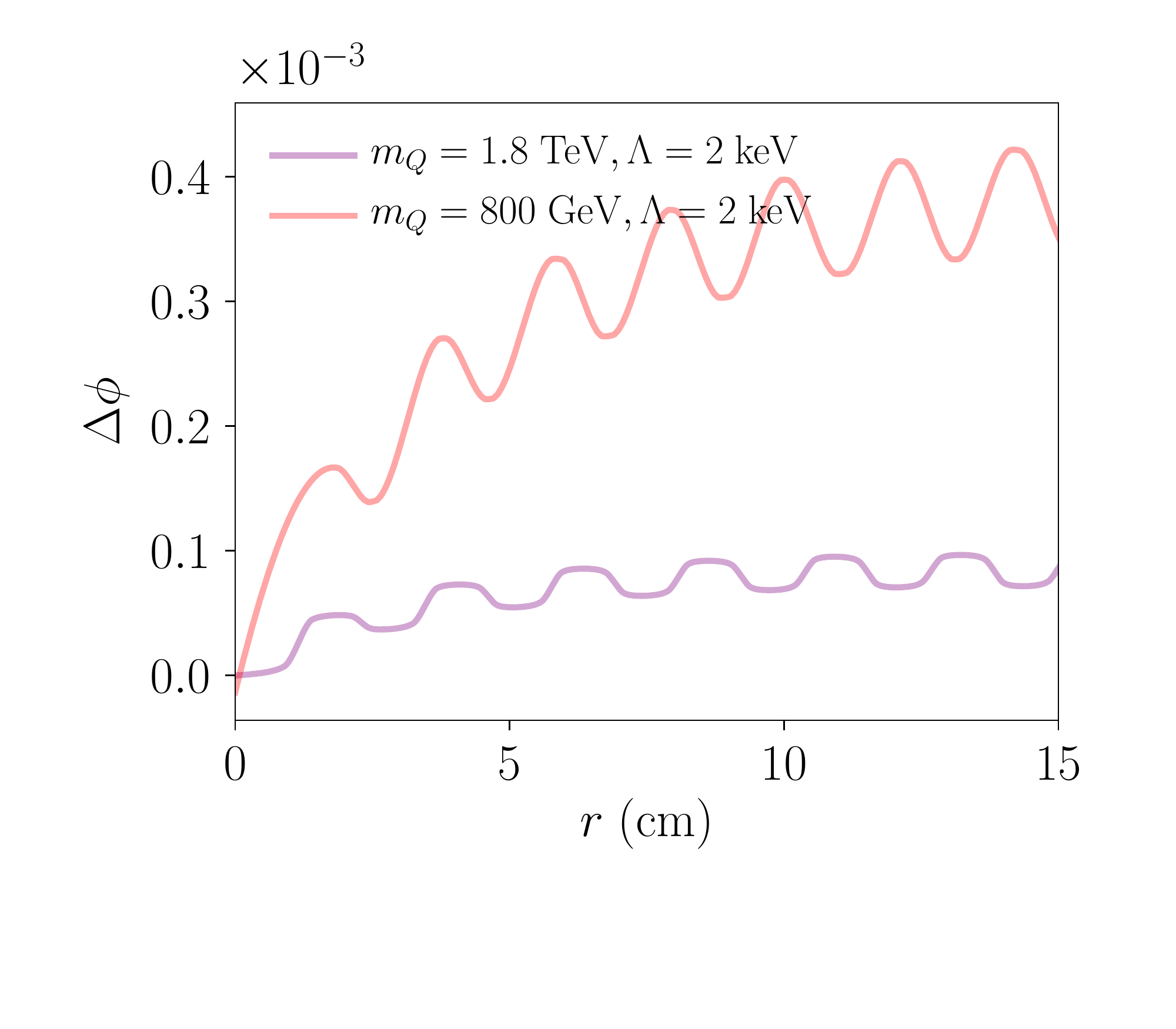}
\caption{$\Delta\theta$ as a function of the radial distance of the center of mass to the origin for two representative benchmark events, with $B=2$ T.}
\label{fig:steps}
\end{figure}

While the internal angular momentum of the system is typically very large in units of $\hbar$, it is still small compared to the resolution of the trackers, and the trajectories remain co-planar as far as the experiments are concerned. We can see this by estimating the angular rotation of the plane spanned by the quirks' velocity vectors:
\begin{align}
\Delta \phi &\sim |{\bm \tau}|  \Delta t^2/ I\sim e v_{\cm}\gamma_\cm\frac{\gamma^2 v^2}{(\gamma-1)^2} \frac{ B}{   \Lambda^2}\\
&\sim  10^{-5}\, \frac{v_{\cm}}{v^2} \,\left(\frac{B}{2\, \mathrm{T}}\right) \left(\frac{2\,\mathrm{keV}}{\Lambda}\right)^2\, .
\end{align}
The key point here is that the effect of the torque on the angular acceleration is suppressed by the large moment of inertia of the system $I\sim 2d^2 m_Q$. There could be an enhancement for close to threshold quirks, where $v\ll 1$;  but this is relevant only for a very small part of phase space, and $\Delta \phi$ is typically not larger than $10^{-3}$.  We show $\Delta \phi$ in Fig.~\ref{fig:steps} for two example events, as found in the full numerical solution of Eq.~\ref{eq:eom} with ${\bm F}_\ext$ the Lorentz force. In the numerical result, $\Delta \phi$ oscillates and slowly accumulates as the quirks travel through the detector until it stabilizes around a fixed value. We see that the typical $\Delta \phi$ is somewhat larger than $10^{-5}$, but is still small compared to the resolution of the tracker. The effect of the magnetic field is accounted for in all our simulations, and any potential efficiency loss due to shifting of the quirks' plane is included in our results. 

Similarly, one can show that the rotation induced by the torque exerted by interactions with the detector material is small: the forces exerted by the ionization process when the quirks traverse a silicon layer are of the order $\sim (100\, \mathrm{eV})^2$, which induce an angular acceleration of up to $\alpha \sim 10\, \mathrm{ns}^{-2}$. The time it takes to traverse a $\sim$ cm thick layer of detector material is $\sim 10^{-2}$ ns, such that the shift in angle is $\Delta \phi\sim10^{-3}$ for each layer the quirks traverse. We therefore neglect this effect in our simulations. It is worth noting that while we focused on the quirk action in Eq.~\ref{eq:nambugoto}, all arguments presented above hold for an arbitrary central force, as long as the external forces are small compared to the central force between the particles.

Finally, a priori dark glueball radiation may also induce a change in angular momentum and therefore, a deviation from co-planarity.  While there is no reliable calculation of the nonperturbative dark glueball radiation rate, naive dimensional analysis suggests that it is irrelevantly small \cite{Kang:2008ea}. Concretely, at large distance, the quirks' glueball radiation rate is proportional to the acceleration, $a\sim \Lambda^2/m_Q$, which is very small compared to the glueball mass $\sim \Lambda$ \cite{Morningstar:1999rf}. This small acceleration strongly suppresses glueball radiation at large distances. On the other hand, when the quirks approach each other within a distance of $\Lambda^{-1}$ or less, $\sim \Lambda$ worth of energy may be radiated in a few glueballs. Such a radiation pattern changes the angular momenta of the quirk/anti-quirk system by a few quanta of $\hbar$, but does not modify the macroscopic trajectory of the quirks.

\section{Search strategy}
\label{sec:strategy}

\subsection{Signal simulation\label{sec:simulation}}

We generate signal samples of vector-like fermions with up to 1 jet using \texttt{MadGraph5\_aMC@NLO}~\cite{Alwall:2014hca, Buchkremer:2013bha}, which is subsequently matched using \texttt{Pythia8}~\cite{Sjostrand:2006za,Sjostrand:2014zea}. For electroweak production the quirks are taken to have unit charge and are produced in Drell-Yan, while for colored production only QCD contributions are included. The resulting four-momenta of the quirks are then evolved by numerically solving the equation of motion in \eqref{eq:eom} assuming a uniform 2T magnetic field along the $z$-direction. The intersections of the trajectories with a simplified model of the ATLAS inner detector are calculated, and the center of each pixel hit is used as the input for our analysis. We hereby use the parameters of the various detector elements as specified in \cite{ATL-PHYS-PUB-2009-002}. Specifically, for the pixel detector we use the pixel size rather than the resolution and for the silicon microstrip tracker (SCT) we conservatively assume a resolution of twice the intrinsic accuracy quoted in \cite{ATL-PHYS-PUB-2009-002}. Hits in neightboring pixels, according to the above definitions, are merged to a single hit, in an attempt to model the ATLAS hit merging procedure. We further apply a uniform, gaussian smearing with width 45 mm to the $z$-coordinates of all the hits, to account for the finite longitudinal size of the beamspot. For simplicity, we only included the barrel of the pixel and SCT detectors in our simulations, which effectively restricts the fiducial range to $|\eta|\lesssim1.8$. Including additional detector components such as the endcaps, calorimeters and/or the transition radiation tracker would further enhance the sensitivity, although it would require a more careful consideration as our co-planar approximation may be invalid for denser materials, and the timing constraint ($t < 25 \, $ns) may become important for components farther away from the interaction point. 





\subsection{Trigger\label{sec:trigger}}
Similar to conventional Heavy Stable Charged Particles (HSCPs), we do not expect quirks with a moderate boost to stop in the material of the calorimeters. This implies that the quirks will typically leave a handful of hits in the ATLAS muon detectors, which may be a triggering opportunity. In particular, the L1 trigger selection requires a coincidence of hits in two or three layers of the muon system, depending on the $p_T$ threshold associated with the trigger path \cite{Aad:2014sca,ATL-DAQ-PUB-2016-001}. The High Level Trigger (HLT) subsequently attempts to reconstruct a track, which is matched to a track in the inner detector. This step is likely to fail for the quirk signature, since a fit to a helix-shaped track is likely very poor for the string tensions we consider here \cite{Farina:2017cts}.  It is however plausible that many of these events could be recovered with a dedicated quirk trigger at the HLT, for example by requiring pairs of nearby hits in multiple layers of the muon system. An important caveat here is that the quirks must reach the muon chamber in less than 25 ns, which may not be the case for a sizable fraction of the events.

If the event contains a sizable amount of transverse energy in the form of initial state radiation (ISR), the HLT will interpret the lack of a reconstructed track as missing transverse energy ($\MET$). With start-up trigger thresholds for run-2 in mind \cite{ATL-DAQ-PUB-2016-001}, we therefore impose a $p_T>$ 200 GeV cut on the center mass momentum of the quirk/anti-quirk system. This requirement implies that the quirk/anti-quirk pair is essentially always central and sufficiently boosted, such that each quirk will most likely intersect each layer only once. The $\MET$ cut also reduces the initial opening angle of the quirk pair, and therefore biases the sample towards smaller oscillation amplitudes. While we will make use of the latter feature, the precise value of the $\MET$ cut does not significantly impact the reconstruction efficiency of our proposed algorithm.

Although the $\MET$ trigger path is conceptually simple, it has a number of important downsides. Firstly, quirks with lower boost can traverse each layer multiple times, which can potentially lead to spectacular events with $\mathcal{O}(10)$ hits in each tracker layer. The requirement of a hard ISR jet removes essentially all of these events.\footnote{It would be interesting to investigate whether some of these events could be recovered with the future CMS and/or ATLAS hardware track triggers \cite{Shochet:1552953,Contardo:2020886}, perhaps along the lines of \cite{Gershtein:2017tsv}.} Secondly, a tight ISR requirement substantially reduces the unusable signal cross sector, which can be problematic especially for Drell-Yan production. Finally, the thresholds for the $\MET$ triggers are expected to increase further as the instantaneous luminosity increases, which will further reduce the signal efficiency. Given that a substantial fraction of the quirk events would likely pass the L1 muon trigger, it would therefore be very interesting to design a suitable trigger path at the HLT which makes use of the muon chambers. Since the focus of this letter is on the off-line reconstruction strategy, we do not consider a potential muon trigger here. 

\subsection{Plane finding Algorithm}
\label{sec:algorithm}
As argued above, the quirk trajectories largely lie on a single plane, which will be the essential ingredient for our proposed algorithm. We will assume that the primary vertex is identified correctly, and is located at the origin. A single hit is then defined by its position three-vector, and a candidate plane is fully specified by its normal unit vector. Our tracking algorithm is thus reduced to solving the following problem: Given a list of hits,  what is the optimal plane that is close to as many hits as possible? To find a solution, one must first define a metric that specifies what `closeness' means. One also needs to define when a hit is considered to be part of a plane, given the finite resolution of the tracker. Finally,  the notion of an `optimal plane' is ambiguous, given that one must weigh the goodness of the fit against the number of hits included.  We will address these issues step by step in the remainder of this section.


A natural choice for the distance measure between a set of hits $\{{\bm x}_a\}_{a\le N}$ and a plane with normal vector ${\bm n}$ is the root-mean-squared distance of the hits to the plane
\begin{align}
  d({\bm n},  {\bm x}_a)\equiv \sqrt{\frac{1}{N-1}\sum^N_{a=1} \left({\bm n}\cdot {\bm x}_a\right)^2}\,.
\label{eq:track_dist}
\end{align}
The distance can be rewritten as $d= \sqrt{{\bm T}_{ij}{\bm n}_i{\bm n}_j}$, where the two-tensor ${\bm T}_{ij}$ is defined by
\begin{align}
  {\bm T}({\bm x}_a)_{ij} \equiv \frac{1}{N-1}\sum_{a=1}^N {\bm x}^a_{i} \,{\bm x}^a_{j}\,.
\label{eq:track_tensor}
\end{align}
Minimization of $d$ with respect to $\bm n$ simply reduces to solving an eigenvalue problem for $\bm T$. The smallest eigenvalue, $\Delta s^2$, then gives the minimum value of $d^2$, with an associated eigenvector ${\bm n}_1$ equal to the normal vector of the optimal plane. $\Delta s$ therefore gives a measure of the thickness of the plane.

There is additional useful information in the other eigenvalues and eigenvectors of ${\bm T}$ that describe the geometry of the hits: Since ${\bm T}$ is symmetric, the eigenvectors are orthogonal. The eigenvectors ${\bm n}_2$ and ${\bm n}_3$, ordered by increasing eigenvalues, therefore lie on the plane defined by ${\bm n}_1$. Geometrically, ${\bm n}_2$ describes a second plane, orthogonal to the first, that has minimal root-mean-squared distance to all the hits. For a pair of quirks on a plane specified by ${\bm n}_1$, the ${\bm n}_2$ plane roughly splits the pair of the hits. Then second eigenvalue, denoted by $\Delta w$, then provide a measure of the width of the quirks' oscillations. As for the third eigenvector ${\bm n}_3$, it is orthogonal to ${\bm n}_{1,2}$ and therefore provides a good estimate of the direction of the quirks' motion. In the limit that $\Delta w$ is small compared to the detector size, all the quirks' hits will then be confined along a narrow planar strip. Specifically, the quirks signal we are after will lie in a positive direction $({\bm x}_a \cdot{\bm n}_3 ) >0$, with a small thickness $\Delta s$ for the fitted plane and an oscillation width $\Delta w$. 

\begin{figure}[t]
\centering
\includegraphics[width=0.45\textwidth, trim={1.3cm 3.3cm 1.3cm .6cm},clip]{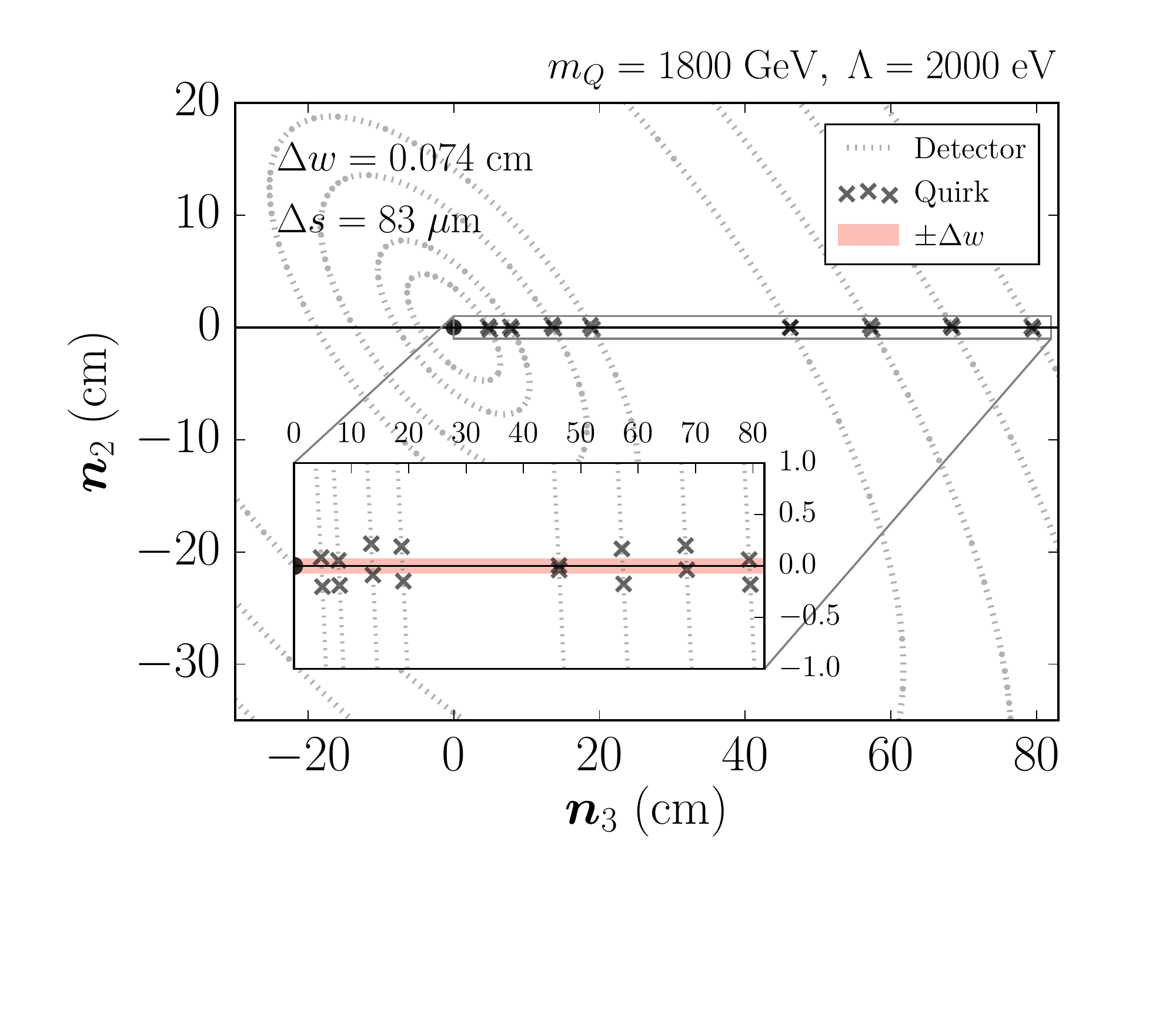}
\caption{Hits for a sample signal event, projected onto the reconstructed plane spanned by $({\bm n}_3,{\bm n}_2)$.  The dotted line shows the cylindrical detector layers projected onto the signal plane as ellipses. The inner figure shows a zoomed-in view of the hit patterns, which lie roughly on a strip with width $\Delta w\sim 0.074$ cm. }
\label{fig:signal_hits}
\end{figure}

Fig.~\ref{fig:signal_hits} shows an example signal hit pattern, projected on the reconstructed plane spanned by $({\bm n}_3,{\bm n}_2)$. The dotted ellipses show the tracking layers projected on the $({\bm n}_3,{\bm n}_2)$ plane. We see that all the hits lay in the positive ${\bm n}_3$ direction, and that the hits mainly lay a few factors within $\Delta w$. As expected, ${\bm n}_3$ reconstructs the quirks' direction to a good approximation.

With the key geometric variables defined, we now describe an algorithm that will iteratively reconstruct an `optimal plane'. Given that for each list of hits, a best fitted plane can be computed as described above, the goal would then be to pick out an `optimal list' of hits $\{{\bm x}_a\}$ among thousands of unassociated hits in an event. The definition of what is optimal will involve a combination of $\Delta s$ and $\Delta w$ cuts, in addition to a few other selection cuts in the algorithm. For simplicity, we assume a detector geometry of 8 layers of detector, following the ATLAS pixel layers and SCT; although our description may be generalized to other detector elements. The algorithm is split into two main stages, the seeding and iterative fitting stage:

\begin{enumerate}
\item Seeding: Define initial hits for iterative fitting
\begin{enumerate}
\item Start from the 8$^{\rm th}$ layer, collect all pairs of hits with $\Delta \phi < 0.1$ and $\Delta z < 2 $cm. Repeat the same for the 7$^{\rm th}$ layer.
\item Construct four-hits combinations by choosing one pair from each initial layer. Compute the tensor ${\bm T}$ and apply the follow cuts: ${{\bm x}_a \cdot {\bm n}_3 >0}$ for all hits, ${\Delta s < 0.05 \;{\rm cm}}$ and ${\Delta w < 1\;{\rm cm}}$.
\end{enumerate}
\item Iterative fitting: for each seed, loop over the remaining 6 layers outside-in, and collect more hits consistent with the initial fit
  \begin{enumerate}
  \item \label{step:it} Start from the 6$^{\rm th}$ layer, collect all hits that satisfy $({\bm x} \cdot {\bm n}_3) > 0$,  ${|{\bm x}\cdot {\bm n}_1| < 0.05\; {\rm cm}}$ and ${|{\bm x}\cdot {\bm n}_2| < 1\; {\rm cm}}$.
  \item Loop over selected hits, starting with the one with the smallest $|{\bm x}\cdot{\bm n}_1|$. Together with the list $\{{\bm x}_a\}$, recompute ${\bm T}$ and associated variables. If $\Delta s$ and $\Delta w$ do not increase by a factor of 3, add the hit to the list.
  \item Iterate the previous steps all the way to the first layer, then construct the final list $\{ {\bm x}_a\}$ and compute associated variables.
  \end{enumerate}
\item Event Selection: Gather all the reconstructed lists, apply the final cut $\Delta w < 1$ cm. If there are more than one plane identified, keep the one with the smallest $\Delta s$.
\end{enumerate}

In summary, after the plane-finding algorithm has identified a set of candidate plains, the main discriminating variables of our analysis are
\begin{itemize}\setlength\itemsep{-0.3em}
\item[--] First eigenvalue of \eqref{eq:track_tensor}, or the ``thickness'', $(\Delta s)^2$
\item[--] Second eigenvalue of \eqref{eq:track_tensor}, or the ``width'', $(\Delta w)^2$
\item[--] Number of hits found
\end{itemize}
It is important to note that the selection cuts on these variables can be easily modified to accommodate better signal acceptance and/or background rejection. A tighter selection will generally boost computational efficiency at a cost of reduced signal efficiency, which is what has been chosen in this work. Looser selection can easily be implemented at a cost of increased computational time, and may require additional adjustments on the final cuts to maintain the same level of background rejection.

An amortized $\mathcal{O}(N)$ time complexity can be achieved for the tracking algorithm, assuming that the $(\Delta \phi, \Delta z)$ cut is adjusted so that roughly a constant number of hits are within such a window.\footnote{Assuming that the hits are stored in such a way that access through $(\phi,z)$ coordinates takes constant time.}  An algorithm of this sort may be sufficiently fast for implementation at the high level trigger (HLT), which would partially remedy the problem of the stringent $\MET$ trigger.

There are additional variables that can potentially enhance background rejection and/or the efficiency of the seeding step. For instance, ${\bm n}_3$ is expected to be aligned with $\,\vMET$ in the transverse plane, which can limit the region of interest in the detector for reconstruction. Additionally, we did not include $dE/dx$ information, which can be leveraged for heavier masses; although we found that the algorithm described above already provided sufficient discriminating power (see Sec.~\ref{sec:results}). Since $dE/dx$ information could nevertheless be of interest for a realistic experimental implementation, we include a brief summary of our relevant results in App.~\ref{app:dedx}.

\subsection{Backgrounds}

The biggest background for our search are unassociated hits, which predominantly come from pile-up tracks for which the track reconstruction failed. For this purpose we use the public available tracking efficiency plots \cite{ATL-PHYS-PUB-2015-051}, where we neglect the $\eta$-dependence of the efficiency, as long as the track is within the $\eta$-range of the barrel. For our study we assume an average of $\langle \mu\rangle=50$ pile-up interactions with a longitudinal beam spot spread of 45 mm, where the former is conservative compared to current conditions by roughly a factor of 2. We model pile-up by randomly selecting minimum bias events from a sample of 125$\times 10^3$ events generated by \texttt{Pythia8}, processed by the simplified detector described in \cite{Knapen:2016hky}. 
We approximately account for all elements of the inner detector, including the beam pipe, service layers and endcaps and include the effects of bremsstrahlung and energy loss from ionization. For more details we refer to appendix A of \cite{Knapen:2016hky}. We did not attempt to model secondaries from hadronic interactions with the inner detector material, which will increase the hit counts in the outer layers of the pixel and SCT detectors. We however verified that this deficiency is roughly offset by our conservative choice for $\langle \mu\rangle$.

A second potential background arises from isolated photon conversions in the beam pipe. These conversions give rise to a fairly collimated $e^+ e^-$ pair, which results in a nearby pair of hits in each layer of the tracker. For some conversion events, these hits could all approximately lie on a plane, and thus fake a quirk signal. We model this background by generating a $Z+\gamma+j$ sample with \texttt{MadGraph5\_aMC@NLO}, where we decay the $Z$ to neutrinos and require at least 200 GeV of $\MET$, to satisfy our trigger requirement. We further require the $p_T$ of the photon be larger than 0.5 GeV. The fiducial cross section for this process is $\sim 1$ pb, which drops to $\sim 10$ fb if we require that the photon converts in the beampipe using the conversion probability from figures 33.16 and 33.17 of \cite{Olive:2016xmw}. Then we assume equal energy sharing between both electrons, which is conservative, as softer electrons would bend more strongly and lead to poor fit to a plane. The $e^+e^-$ pair is then passed through the same detector simulation as described for the pile-up background. We subsequently overlay pile-up hits and pass the resulting set of hits through our reconstruction algorithm. 


\section{Results\label{sec:results}}

\begin{figure*}[t!]
\begin{minipage}{.47\textwidth}
\includegraphics[trim={0.0cm 3.cm 0cm 1cm},clip,width=1.0\linewidth]{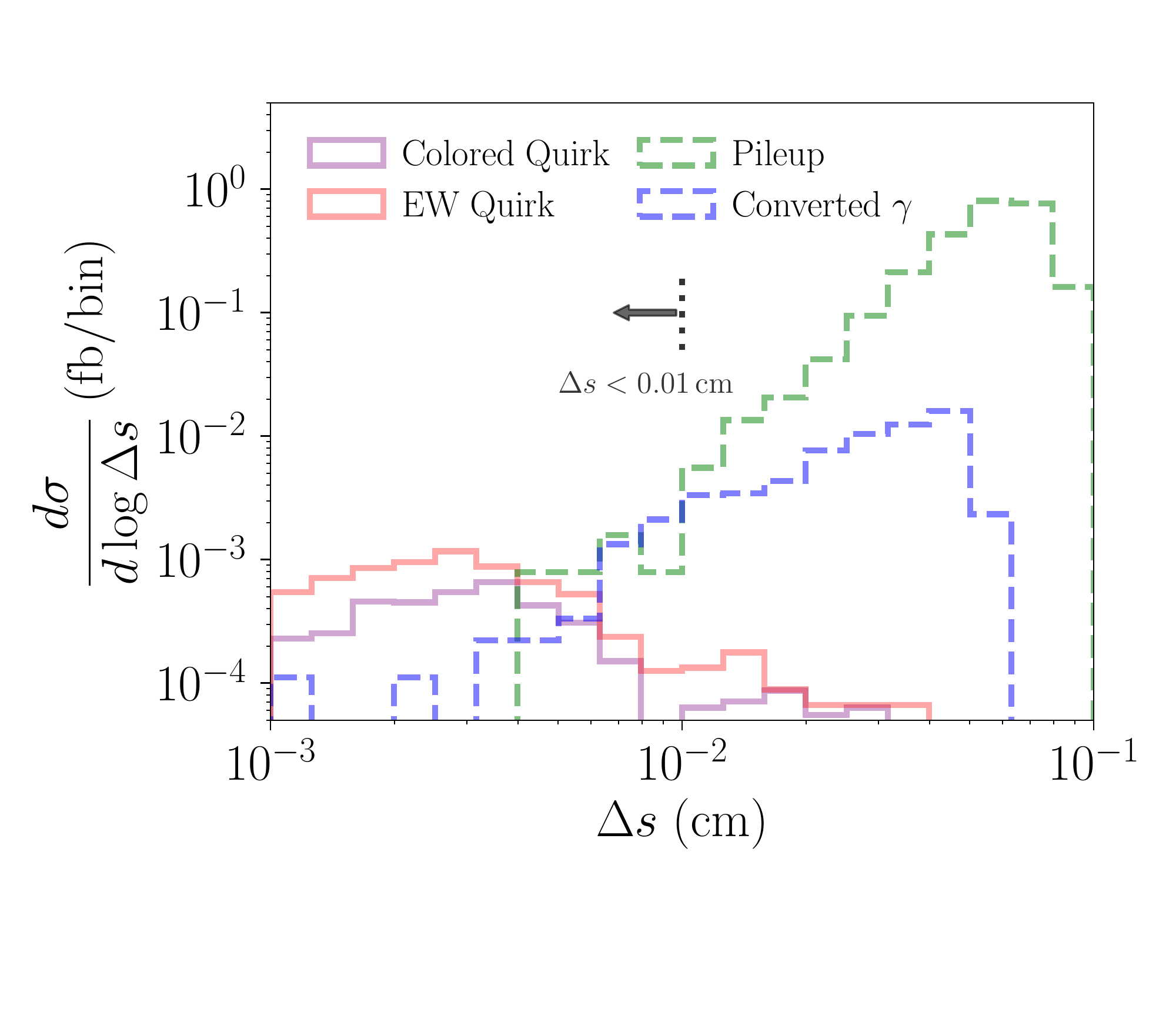}
\end{minipage}
\hspace{.3cm}
\begin{minipage}{.47\textwidth}
\includegraphics[trim={0.0cm 3.cm 0cm 1cm},clip,width=1.0\linewidth]{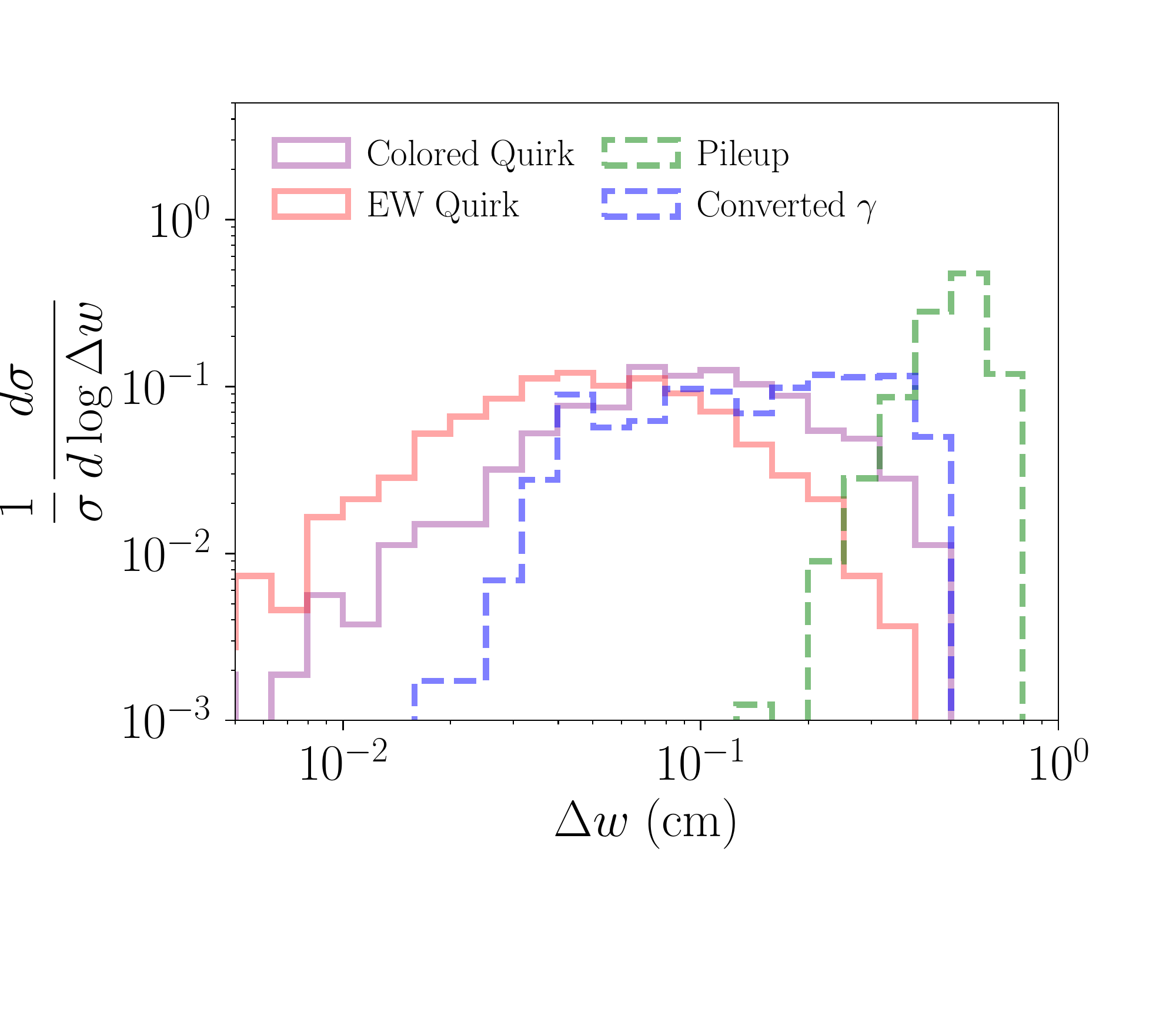}
\end{minipage}
\caption{Signal and background distributions for thickness and width of the strip, parametrized by $\Delta s$ (left) and $\Delta w$ (right) respectively. The signal benchmarks for colored (EW) quirks are given by $m_Q = 1.8\; {\rm TeV}$ and $\Lambda=2000\;{\rm eV}$ ($m_Q = 800\; {\rm GeV}$ and $\Lambda=4000\;{\rm eV}$). A signal selection cut $\Delta s < 0.01$ cm is indicated on the left figure. }
\label{fig:variables}
\end{figure*}

Given the $\mathcal{O}(1000)$ unassociated pile-up hits per layer in the tracker, a subset of these hits do accidentally land on a plane. Through our reconstruction algorithm, only $\sim 10^{-3}$ of all background events contain a plane with at least one hit in 4 out of 8 layers. The number rapidly drops to $6\times 10^{-5}$, for events with a plane that contains at least one hit in each layer. Still the majority of these planes have only one hit in most of the layers. Our signal region is then defined by the following cuts:
\begin{enumerate}
\item At least one plane reconstructed under the tracking algorithm
\item \label{cuts:hits} All but one layer must contain 2 hits, the remaining layer must contain at least 1 hit
\item ${\Delta w < 1.0 \; {\rm cm}}$ and ${\Delta s < 0.01 \; {\rm cm}}$ 
\end{enumerate}
For a quirk signal, as long as the length scale of oscillation $m_Q/\Lambda^2$ is smaller than $\sim 1\;{\rm cm}$, and if the quirks pass through all 8-layers, the reconstruction efficiency for these cuts can be as high as $\sim 73\%$.  

Fig~\ref{fig:variables} shows the $\Delta s$ and $\Delta w$ distribution for background and our benchmark signal before the final cut on those variables are imposed. We see that the signal and background have distinctive distributions. In order to compensate for the lack of simulation statistics, the pileup backgrounds are derived by taking the distribution from events that are allowed to have 1-hit per layer, weighted by the overall efficiency of passing the more stringent requirement in point \ref{cuts:hits} above. For the pile-up background, the number of hits is anti-correlated with the thickness and the width of the plane, and as such this yields a conservative estimate for the pile-up background in the signal region. We deliberately do not impose a tight cut on $\Delta w$, as the efficiency for such a cut is strongly signal dependent. The rather loose cut of ${\Delta w < 1.0 \; {\rm cm}}$ is intended to retain decent efficiencies for quirks with larger oscillation amplitude (low $\Lambda$). Even though Fig~\ref{fig:variables} suggests a few background events after the final selection cut of $\Delta s < 0.01$ cm, we suspect that they can easily be removed through either a $\Delta \phi$ requirement between $\vMET$ and ${\bm n}_3$, and/or by examining $dE/dx$ pattern for the reconstructed hits. We have also not used any direct information on the quirk trajectory, other than the semi-strip geometry. Should our background estimates prove to be overly optimistic in a real experimental setup, it should be possible to further increase signal discrimination by fitting a quirk trajectory to the hits identified by our method. If any quirk candidates are observed, this would also be an obvious way to try to measure the mass and string tension.

We factorize the total signal efficiency into the trigger efficiency ($\epsilon_{\trig}$), the fiducial efficiency ($\epsilon_{\fid}$) and the reconstruction efficiency ($\epsilon_{\reco}$) such that the total efficiency $\epsilon$ is given by
\begin{equation}
\epsilon = \epsilon_{\trig}\times  \epsilon_{\fid}\times  \epsilon_{\reco}\,.
\end{equation}
The trigger efficiency tends to be low, especially for Drell-Yan production, but it may be possible to improve on this with dedicated trigger strategies, as outlined in Sec.~\ref{sec:trigger}. The fiducial efficiency parametrizes the likelihood that each quirk intersect with each layer at least once, in events passing the trigger. We also include a 25 ns timing cut, which causes a slight drop in $\epsilon_{\fid}$ for heavier quirks, which tend to be slower. Inclusion of the endcaps should increase $\epsilon_{\fid}$ without significantly impacting the tracking algorithm. The reconstruction efficiency is defined as the efficiency of our algorithm in finding quirks which satisfy both the trigger and fiducial requirements. The various efficiencies are shown in Tab.~\ref{tab:efficiencies} for two benchmark points. We see that the peak $\epsilon_\reco$ can be as high as $\sim 70\%$, while $\epsilon_\reco$  drops at lower $\Lambda$, where the iterative algorithm may fail to capture enough hits largely due to a stringent $\Delta w$ requirements. At high $\Lambda$, $\epsilon_\reco$ drops as well since the separation is small enough for the hits to start merging, at which point a plane cannot be found. 


\begin{table}[b]
\begin{tabular}{|c|c|ccc|}\hline
$m_Q$ (GeV) &$\Lambda$ (keV)&$\epsilon_{\trig}$&$\epsilon_{\fid}$&$\epsilon_{\reco}$\\\hline
\multirow{5}{*}{\begin{tabular}{c}$800$\\(DY)\end{tabular}}&1&\multirow{6}{*}{0.10}&\multirow{6}{*}{0.28}&0.11\\
&2&&&0.41\\
&3&&&0.65\\
&4&&&0.72\\
&5&&&0.74\\
\hline
\multirow{6}{*}{\begin{tabular}{c}$1800$\\(QCD)\end{tabular}}&1&\multirow{6}{*}{0.24}&\multirow{6}{*}{0.28}&0.083\\
&2&&&0.35\\
&3&&&0.59\\
&5&&&0.74\\
&10&&&0.58\\
\hline
\end{tabular}
\caption{Breakdown of the signal efficiencies for two benchmarks, one for Drell-Yan (DY) production, and one for colored production (QCD). $\epsilon_{\trig}$ and $\epsilon_{\fid}$ are independent of $\Lambda$, the latter with the exception of small edge effects. For $\Lambda\gtrsim 5$ keV, $\epsilon_{\reco}$ deteriorates as pairs of hits start merging into a single pixel.}
\label{tab:efficiencies}
\end{table}

\begin{figure*}[t]
\begin{minipage}{.47\textwidth}
\includegraphics[trim={0.1cm 3.cm 0cm .5cm},clip,width=1.0\linewidth]{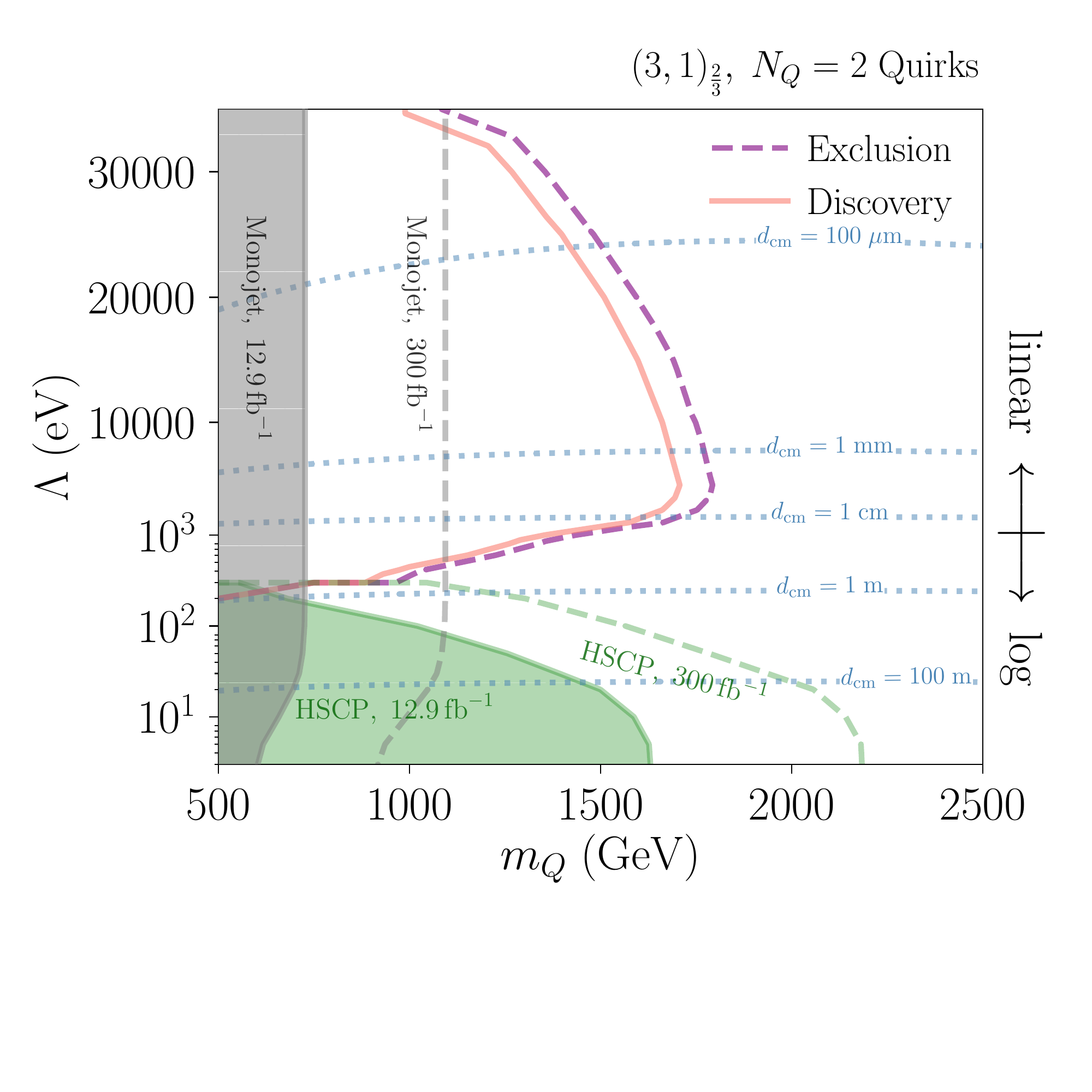}
\end{minipage}
\hspace{.3cm}
\begin{minipage}{.47\textwidth}
\includegraphics[trim={0.1cm 3.cm 0cm .5cm},clip,width=1.0\linewidth]{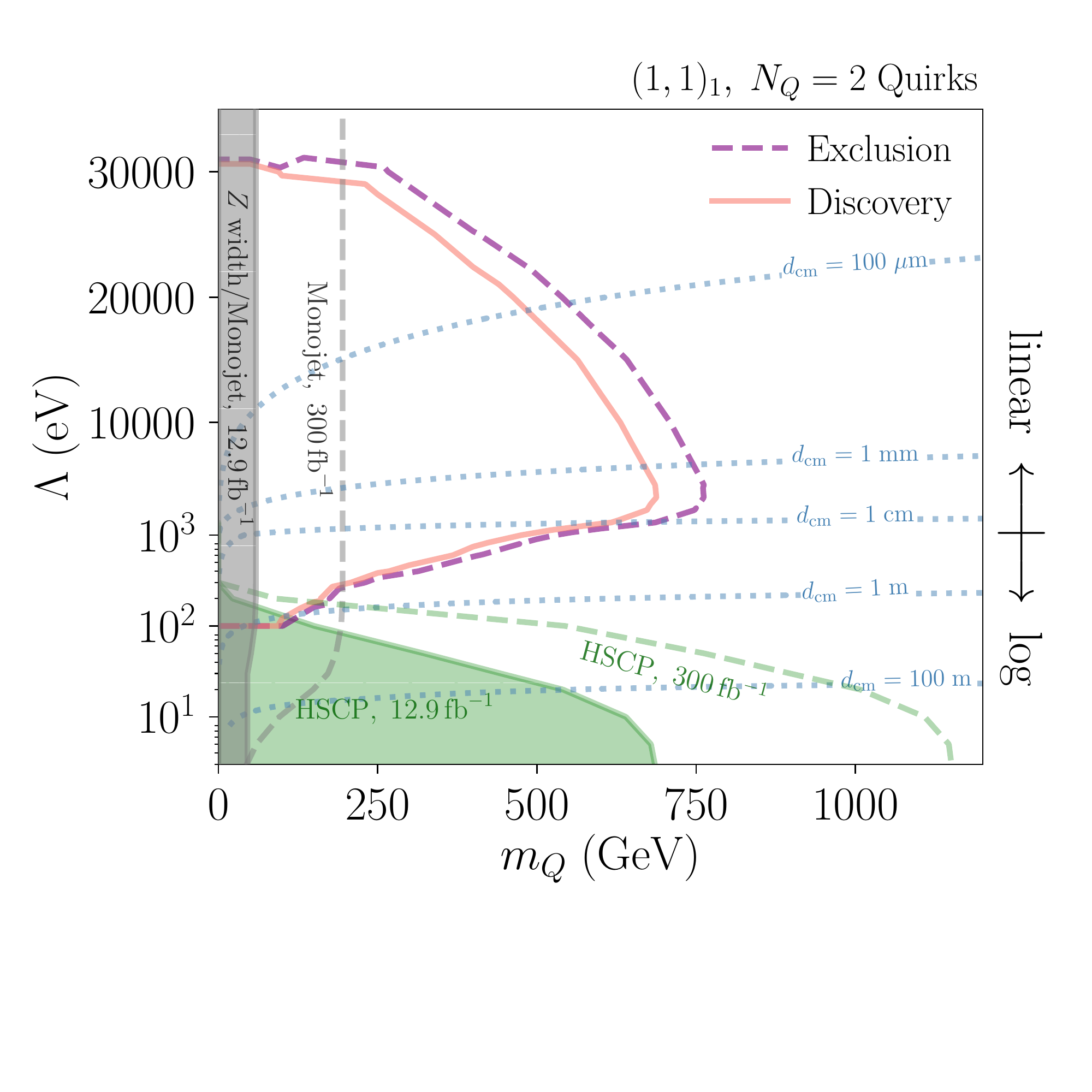}
\end{minipage}
\caption{Contours of having 3 (5) events in the $m_Q$ vs $\Lambda$  plane for an integrated luminosity of $\int L dt = 300\;{\rm fb}^{-1}$, overlaid with (projected) HSCP and monojet limits \cite{hscplimit}, where we extrapolated the latter to high $\Lambda$. In reality, the monojet limits may deteriorate in high $\Lambda$ part of the plot, where the quirk system may be reconstructed as a single, high $p_T$ track.  The 3 events bound corresponds to 2-$\sigma$ exclusion given no background. Discovery is defined by 5 events, which may be achieved by close examination of each individual event and by showing that they are compatible with a fixed mass and tension. The dashed blue contour shows the average separation of the quirks in the CM frame, $d_{\rm cm}$, as defined in Eq.~\ref{eq:d_cm}.}
\label{fig:limit}
\end{figure*}

In Fig.~\ref{fig:limit}, we show the expected 95\% exclusion for an integrated luminosity of 300 ${\rm fb}^{-1}$, assuming negligible irreducible backgrounds. We also show a tentative `discovery' curve, corresponding to an expected signal of 5 events. (Discovery with only a few events may be possible when multiple events show hit patterns consistent with a common $(m_Q,\Lambda)$.) Our results are highly complimentary with recent (projected) constraints from HSCP searches \cite{Farina:2017cts}, which are very stringent in the low string tension regime.

In summary, we show that searching for planar hits in the inner tracker is a powerful way to search for quirks with intermediate string tensions. It is moreover possible to design a generic search, which has good acceptance to all string tensions and quirk masses in the qualitative regime of interest ($\mu$m-cm size oscillations). Additionally, we show that an efficient algorithm can be straightforwardly implemented, while providing strong background rejection. While our theory study is no substitute for a full analysis, including understanding more subtle detector effects and backgrounds, we are optimistic that this type of search could be (nearly) free of irreducible backgrounds, especially if a quirk track is fitted to the hits identified by a plane-finding algorithm.




\textbf{Acknowledgements}
We are grateful to Jared Evans, Marco Farina, Maurice Garcia-Sciveres, Laura Jeanty, Matthew Low, Markus Luty, Benjamin Nachman, Simone Pagan Griso and Jesse Thaler for useful discussions. We thank Marco Farina and Matthew Low for supplying us with the HSCP and monojet limits in Fig.~\ref{fig:limit}, and thank Jared Evans and Matthew Low for comments on the manuscript. We further thank Wei-Ming Yao for bringing the conversion background to our attention. 
SK, HL and MP are supported in part by the LDRD program of LBNL under contract DE-AC02-05CH11231, and by the National Science Foundation (NSF) under grants No. PHY-1002399 and PHY-1316783. JS was supported by the Science and Technology Facilities Council (UK). JS is grateful to the Berkeley Center For Theoretical Physics and Lawrence Berkeley National Lab for their hospitality when part of this work was completed. This research used resources of the National Energy Research Scientific Computing Center, which is supported by the Office of Science of the DoE under Contract No.~DE-AC02-05CH11231.

\appendix
\section{$dE/dx$ information\label{app:dedx}}

\begin{figure*}[t]
\begin{minipage}{.45\textwidth}
\includegraphics[trim={0.1cm 2.cm 0cm 1cm},clip,width=\textwidth]{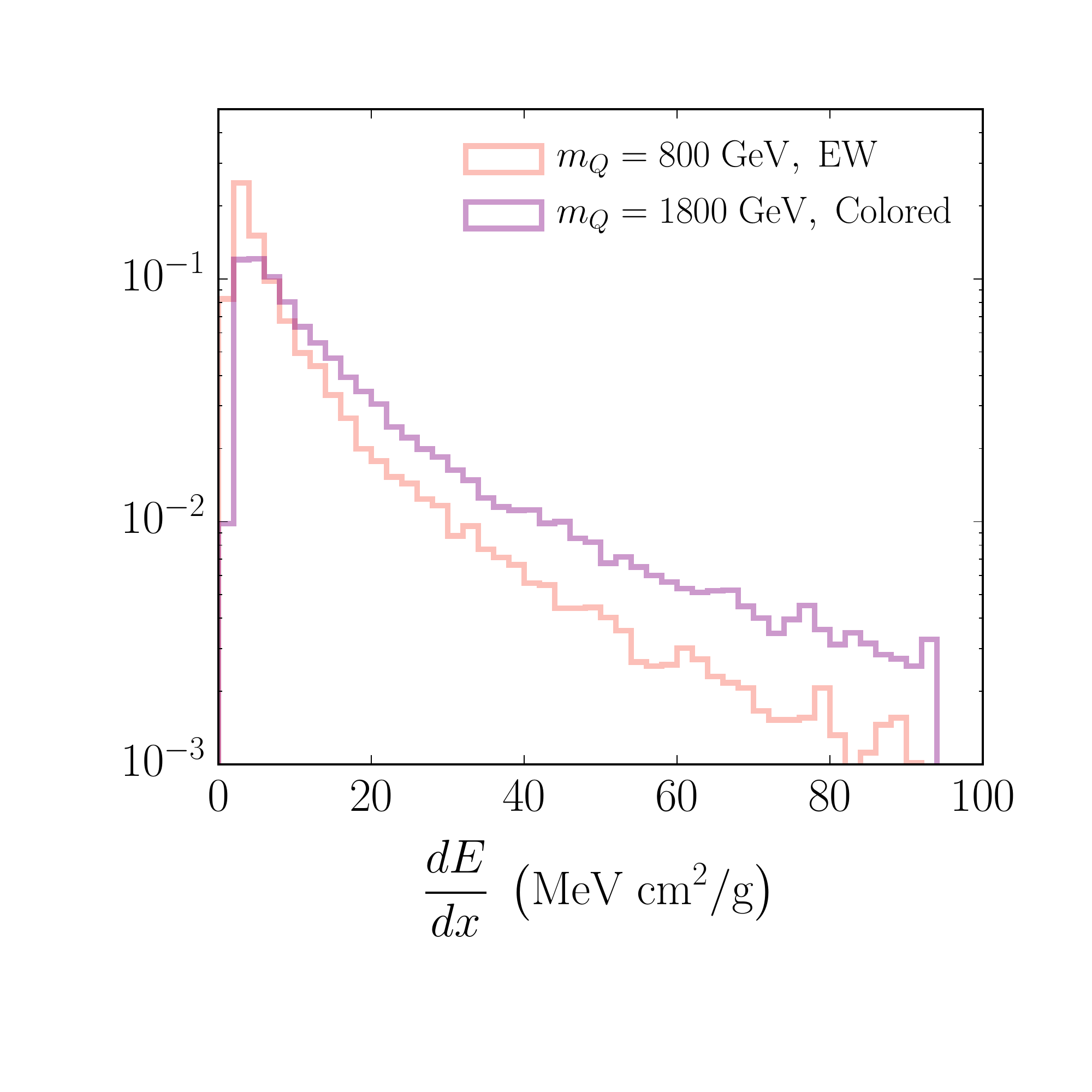}
\end{minipage}
\hspace{.3cm}
\begin{minipage}{.45\textwidth}
\includegraphics[trim={0.1cm 2.cm 0cm 1cm},clip,width=\textwidth]{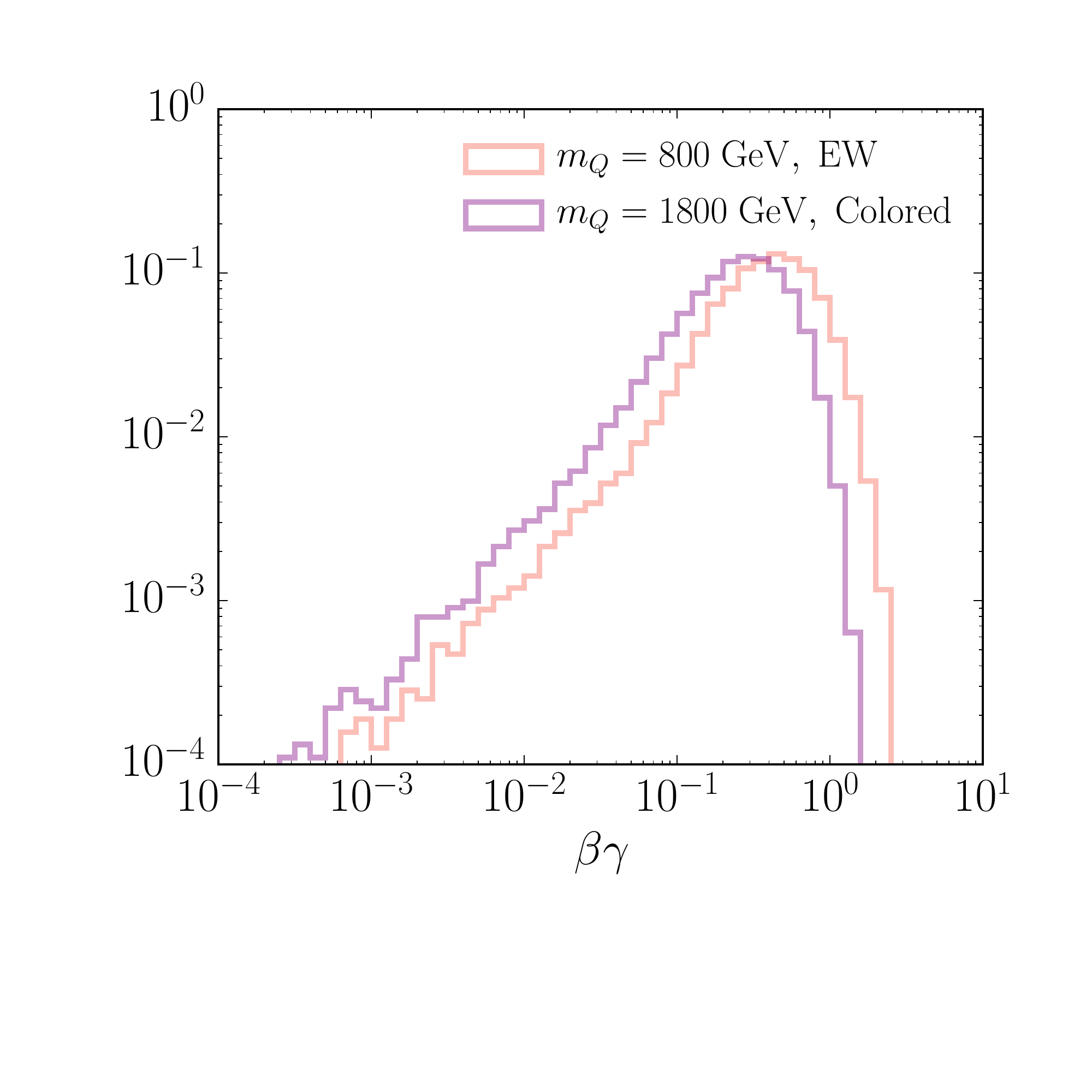}
\end{minipage}
\caption{Differential distribution of $dE/dx$ (left) and $\beta\gamma$ (right) for two signal benchmark points with $\Lambda=2$ keV in our simulation of the ATLAS pixel detector. }
\label{fig:dedx}
\end{figure*}

Although we did not make use of variables relying on $dE/dx$ measurements in our analysis, we here include a brief discussion for completeness. In Fig.~\ref{fig:dedx} we show the include the $dE/dx$ and $\beta\gamma$ distributions for the hits in the ATLAS pixel detector for a few signal benchmarks. For the $dE/dx$ we use the most probable value as a function of $\beta\gamma$ \cite{Olive:2016xmw}. Since this simplified treatment of the $dE/dx$ distribution is not accurate for very slow particles, we omitted hits with $\beta\gamma<0.1$ in the left-hand panel of Fig.~\ref{fig:dedx}. While $dE/dx$ is a powerful variable in conventional HSCP searches, its utility for quirks is more subtle because the quirks accelerate and decelerate along their trajectory through the detector. This implies that a substantial fraction of hits has rather low $dE/dx$, and as such a tight cut is most probably not advisable if a high signal efficiency is desired. On the other hand, should an excess of events be observed, we expect that $dE/dx$ will be important to measure the mass and string tension.

\FloatBarrier

\bibliographystyle{utphys}
\bibliography{quirksbib}

\end{document}